\documentclass[conference]{IEEEtran}
\IEEEoverridecommandlockouts

\usepackage{cite}
\usepackage{amsmath,amssymb,amsfonts}
\usepackage{algorithmic}
\usepackage{graphicx}
\usepackage{textcomp}
\usepackage{xcolor}
\def\BibTeX{{\rm B\kern-.05em{\sc i\kern-.025em b}\kern-.08em
    T\kern-.1667em\lower.7ex\hbox{E}\kern-.125emX}}
\begin{document}

\title{God's Innovation Project - Empowering The Player With Generative AI\\
{\footnotesize \textsuperscript{}}
}

\author{\IEEEauthorblockN{Ritvik Nair}
\IEEEauthorblockA{\textit{Electrical and Computer Engineering} \\
\textit{New York University}\\
New York, United States of America \\
rn2520@nyu.edu}
\and
\IEEEauthorblockN{Timothy Merino}
\IEEEauthorblockA{\textit{Computer Science and Engineering} \\
\textit{New York University}\\
New York, United States of America \\
tm3477@nyu.edu}
\and
\IEEEauthorblockN{Julian Togelius}
\IEEEauthorblockA{\textit{Computer Science and Engineering} \\
\textit{New York University}\\
New York, United States of America \\
julian@togelius.com}
}

\maketitle

\begin{abstract}
In this paper, we present \textit{God’s Innovation Project (GIP)}, a god game where players collect words to dynamically terraform the landscape using generative AI. A god game is a genre where players take on the role of a deity, indirectly influencing Non-Player Characters (NPCs) to perform various tasks. These games typically grant players supernatural abilities, such as terrain manipulation or weather control. Traditional god games rely on predefined environments and mechanics, typically created by a human designer. In contrast, \textit{GIP} allows players to shape the game world procedurally through text-based input. Using a lightweight generative AI model, we create a gamified pipeline which transforms the player's text prompts into playable game terrain in real time. To evaluate the impact of this AI-driven mechanic, we conduct a user study analyzing how players interacted with and experienced the system. Our findings provide insights into player engagement, the effectiveness of AI-generated terrain, and the role of generative AI as an interactive game mechanic.
\end{abstract}


\section{Introduction}
Procedural content generation (PCG) has been used as a core game mechanic since its first usage in video games. Whether it be for map generation in open-ended games like Minecraft, creating maps for a rogue-like, or even generating types of games\cite{b13}.

On the other hand, AI in video games started off as being used to control NPC behavior, creating allies or adversaries for the player. Over time, the usage for AI in games has expanded to other aspects such as game playing and level generation.

The recent rush of improvements in the field of generative AI has brought about the exploration of combining the two techniques into AIPCG. One such method is done through Neural Cellular Automata (NCA)\cite{b10}. It was demonstrated that the NCAs—neural networks that emulate cellular automata behavior—can be used to iteratively construct game levels. This method allows for the emergence of complex patterns from simple, local interactions. Another method that utilizes generative AI for level generation is Path of Destruction \cite{b11}. This approach views level generation as a repair process. Starting with a dataset of existing levels, the method generates an artificial dataset by introducing various sequences of mutations to these levels. 

In this paper, we aim to push AI beyond its conventional role by integrating generative AI as a core game mechanic. Our goal was to explore how players interact with and respond to AI-driven terrain generation. To achieve this, we incorporated the Five Dollar Model, a lightweight text-to-image generator, allowing players to input text that is then translated into terrain modifications in real time.

As part of this study, we also investigate the sentiment a player has towards the integration of generative AI in video games. Recent sentiment towards AI has been negative, especially when it comes to creative fields such as game development and artists. From controversies involving AI art to strikes done by voice actors regarding the use of AI voices in video games and cartoons, many have voiced their opinions against generative AI.

With this in mind, we investigate player reactions to AI-generated terrain as a game mechanic they can interact with. We evaluate whether the system effectively met player expectations and how diverse the resulting landscapes were. To gather insights, we logged player inputs and conducted a user survey, analyzing how players engaged with and perceived the generative AI system in the context of gameplay.

\section{Background}
Procedural content generation (PCG) has been a fundamental aspect of game design for decades, enabling developers to create expansive and dynamic environments without the need for manual level design. Traditionally, PCG techniques rely on algorithms such as Perlin noise, cellular automata, and rule-based systems to generate content such as landscapes, dungeons, and enemy placements. With the rapid advancement of artificial intelligence (AI), researchers have begun exploring the integration of machine learning models into procedural generation, pushing the boundaries of adaptability and player-driven content creation.

The paper Procedural Content Generation via Machine Learning (PCGML) \cite{b9} discusses the potential of AI-driven approaches in PCG. One of the key research challenges outlined in the paper is the use of machine learning models not just for passive content creation but as an active game mechanic—a concept that has remained largely unexplored outside of a handful of games such has \textit{1001 Nights}\cite{b12}.

This paper aims to build upon that idea by demonstrating how generative AI can function as an interactive mechanic, allowing players to influence the game world dynamically through natural language input.

A significant challenge in AI-driven procedural generation is ensuring that the generated content is both coherent and meaningful within the context of gameplay. Unlike traditional PCG, where outputs are constrained by hand-crafted rules, machine learning-based PCG often introduces an element of unpredictability. While this sense of unpredictability can be leveraged to serve the game play loop --- giving players a new experience each playthrough, or a sense of exploration and discovery --- it also lead to frustration. Generated content may result in unplayable or unpredictable artifacts, ruining the experience. This raises the question: How do players perceive AI-generated content, and do current PCGML algorithms align with their expectations? We explore this as a central theme of our research.

One notable example of AI-driven content generation is Sentient Sketchbook \cite{b7}, an AI-assisted level design tool. The tool provides designers with real-time feedback and generates alternative map layouts based on their initial designs. More than just a visualization aid, the system actively evaluates playability and suggests improvements by analyzing structural elements within the game environment. This hybrid approach—where AI serves as a collaborative tool rather than a replacement for human creativity—has inspired our work in \textit{GIP}. However, unlike Sentient Sketchbook, where AI assists the developer, our work focuses on player-driven generation, where AI assists the player in shaping their own gameplay experience.

Generative AI models, particularly those using deep learning architectures, have demonstrated remarkable capabilities in producing creative outputs based on human input. The Five-Dollar Model \cite{b1} is an example of a lightweight model designed for low-dimensional image generation, capable of producing sprites and tile-based maps from textual prompts. The model operates within a 10×10 grid, using a predefined set of 16 tiles to generate visually coherent maps. Unlike high-complexity models such as GANs (Generative Adversarial Networks) or diffusion-based AI models, the Five-Dollar Model prioritizes efficiency and speed, making it well-suited for real-time player interactions within a game.

By integrating the Five-Dollar Model into \textit{GIP}, we aimed to explore its potential as a game mechanic rather than just a content creation tool. In our implementation, players are given a set of collectible words that serve as input for the AI model. The generated terrain is not only a visual effect but directly influences gameplay mechanics, such as movement restrictions, strategic positioning, and resource availability. This interactive AI system allows for emergent gameplay, where different players may develop unique strategies based on the words they collect and the terrain they generate.

By leveraging generative AI as a core gameplay mechanic, our work builds upon existing research in PCG, AI-assisted game design, and real-time content generation. Unlike traditional procedural generation methods, where player influence is minimal, \textit{GIP} empowers players to shape their environment dynamically using AI-driven terraforming. Through our study, we aim to contribute to the growing field of player-AI collaboration in game design, providing insights into how AI-generated mechanics can enhance creativity, strategy, and overall player experience.

\section{Game Overview}
God’s Innovation Project is a god game that revolves around the player interacting with controllable NPCs to collect words and terraform the environment to their advantage.
The goal is to defeat all the monsters that spawn on the map. 

\subsection{Terraforming}
\begin{figure}
    \centering
    \includegraphics[width=0.75\linewidth]{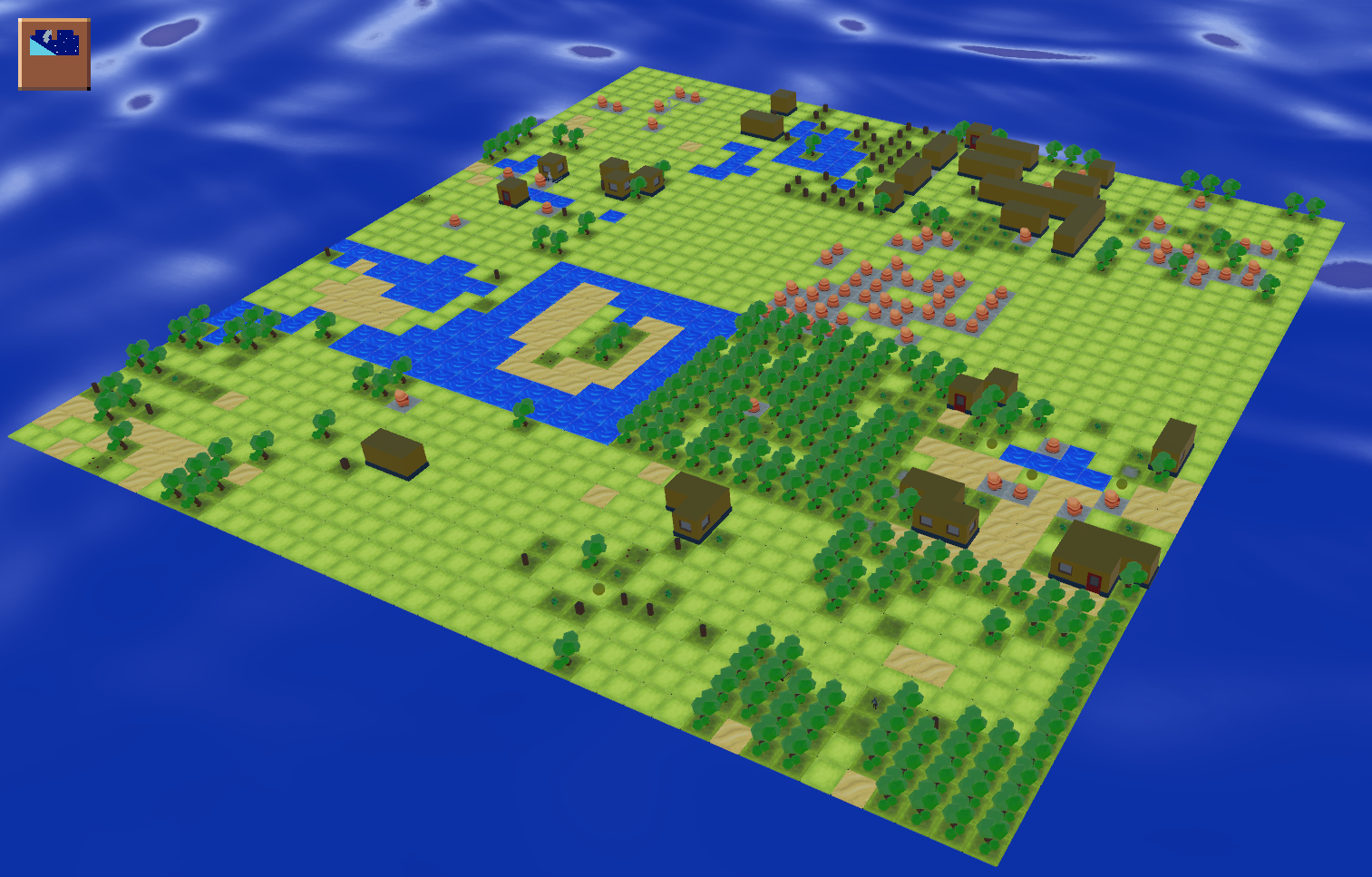}
    \caption{A screenshot of the in-game world. Each world is composed of 16 10x10 grids that can be terraformed by the player.}
    \label{fig:game}
\end{figure}

The core mechanic of the game, and the mechanic that leverages PCGML, is terraforming. The playable game world is divided into sub-grids of ten by ten tiles. Each sub-grid is represented as a ten by ten matrix of integers, ranging from zero to fifteen. These integers each correspond to a tile from an existing tileset. In the current version of the game, the game world is composed of 16 sub-grids, arranged in a four by four square, for a total size of 40 by 40 tiles (Figure \ref{fig:game}).

During gameplay, the player is able to select any sub-grid to terraform by providing a text prompt to the generative model. This prompt is sent to the generative model, hosted on an external server, and the generated map is spawned in at the selected location.

While the generative model allows for natural language input, this modality poses a few issues for use in the core gameplay loop. Players may often choose to enter random or nonsensical prompts to test the limits of the system, which may lead to an early loss. Alternatively, players may identify optimal prompts to use, and repeatedly spawn these map chunks for a trivial victory.

We avoid this issues, and add an interesting new dynamic to the system, by game-ifying the text prompt system. Rather than allowing free-form text input, we restrict the player to compose their prompt using only words in a word bank. We make these words a resource gathered from the game world, which provides incentive for the player to engage with the resource loop as well as a sense of discovery and satisfaction in uncovering new words to use.
\subsubsection{Generative Model}
We utilize the pretrained Five Dollar Model (and accompanying sentence embedding model) using the ``map'' domain. While the original model focuses entirely on aesthetics (ignoring functional gameplay constraints), the ``map'' tileset contains all of the necessary tile types for a functional game. As such, we can directly reuse the trained model by designing the mechanics of GIP around this tileset.

The model is a simple feed-forward convolutional network, which takes a sentence embedding vector as input and returns a 10 by 10 grid of integers corresponding to the tileset. To improve performance and compatibility, we host this model on an external server and serve generated content via simple HTTP requests. This eliminates the need to host the model (or the larger text embedding model) locally, removing some hardware restrictions for players and user study participants.

\subsubsection{Pre-processing}
To construct the word bank, we use the original dataset used to train  The Five Dollar Model \cite{b1}. This restricts the word pool to be fully in-distribution for the model, while allowing new combinations of words to explore the limits of its generative capabilities. We select the top 1000 most common words from the training set to be our word pool.

\subsubsection{Post-processing}
While the original tileset contains most tiles needed for a god game, we incorporate some post-processing to make the world to feel more unique and alive. For example, we modify the rock block based on neighbouring rocks. The more neighbouring rocks, the larger each rock tile becomes. This allows us to create mountains rising from the terrain, rather than being restricted to the original 2D grid world. Similarly, we modify the water tiles such that they connect to nearby water tiles, creating water features like rivers and lakes. This can help create the illusion of water features that are larger than the model natively allows, by connecting tiles spawned in separate sub-grids.

\subsection{Tiles}
\begin{figure}
    \centering
    \includegraphics[width=0.75\linewidth]{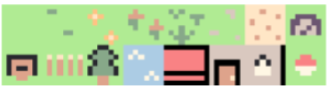}
    \caption{The original 2D map tileset.}
    \label{fig:origtileset}
\end{figure}
\begin{figure}
    \centering
    \includegraphics[width=0.75\linewidth]{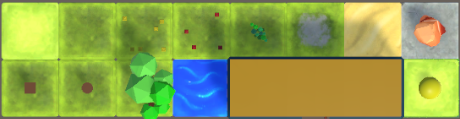}
    \caption{The recreated 3D map tileset.}
    \label{fig:recreatedTileset}
\end{figure}

The game contains 16 different tiles, corresponding to the original ``map'' domain tileset (Figure \ref{fig:origtileset}). We recreated the tiles in 3D (Figure \ref{fig:recreatedTileset}) to use for GIP. For simpler tiles that do not have a significant 3 dimensional aspect, such as grass, sand, path, and water, we simply used an asset that had the appropriate texture for said tiles. For tiles that could be easily recreated in 3D; such as trees, flowers, and bushes; we made 3D models for them and placed said models on top of a grass tile as one tile.

Recreating the house-related tiles posed the most challenge. As the original tileset was meant to be used in a 2D game, it had separate tiles for doors, windows, and roofs. In a 3D game, having a separate tile for roofs does not make sense. Additionally, each of the tiles needed to be able to be put next to each other and still look good despite how many were placed next to each other. Thus, we made all the house tiles into a uniform block that would look good when placed next to each other regardless of how many and shape. In addition, to preserve the uniqueness of each tile, we gave the door tile a door, and the window tile a window.

Each tile comes with a unique set of properties. Taking the example of the tiles mentioned above, the rock tile prevents any NPC from crossing it and water slows the NPCs down. 



\subsection{Words}
Words are a key aspect of our game, and are the mechanism by which the player interacts with the generative model. Words can be obtained through three means. The first (and primary) method is through tree tiles. Any tree tile in the game can be cut using a villager npc. This will grant the player a random word based on a gacha-like probability system. The system takes the top 1000 words and splits them at a 1:9 ratio based on appearance rate in the training dataset. Then, whenever a tree is cut, any word from one of the two word groups is randomly selected based on a 50:50 ratio. This means that a word from the first 100 words is much more likely to be selected than the last 900. Once a word is selected, that word is sent to the back of the queue, and all other words are moved up by one. This allows words near the top of the 900 list to enter the 100 list while moving more common words to the back. This allows the player to experience some unique, uncommon words while still giving them a variety of useful ones.

\begin{table}[h]
    \centering
    \begin{tabular}{lc}
        \textbf{Top 5 Words} \\
        with & 728 \\
        and & 458 \\
        trees & 407 \\
        forest & 266 \\
        path & 256 \\
        \textbf{Bottom 5 Words} & \textbf{Frequency} \\
        rest & 1 \\
        identical & 1 \\
        misaligned & 1 \\
        blobs & 1 \\
        offcenter & 1 \\
    \end{tabular}
    \caption{Top 5 and Bottom 5 Words Used in the Dataset.}
    \label{tab:top_bottom_words}
\end{table}

Another way the player can obtain new words is through ``treasure balls'' that can appear through certain prompts. When collected by a villager, these give the player a random assortment of 5 words (regardless of rarity). Due to the rarity of the treasure ball tile, the player might never encounter it in a playthrough or they might find many if they are lucky.

The final method to obtain words occurs at the very beginning of the game. To provide the player with a starting point from which to collect new words,the player is given the word ``forest''. ``Forest'' is a very powerful word, as the resulting generated map will naturally contain many tree tiles in a concentrated area. Thus, the player is able to obtain a variety of words early in the game.

\subsection{Villagers}
The game contains three NPC types which the player is able to command. These types are: fighter, archer, and worker. The fighter and archer are combat-focused NPCs, which can be tasked to fight enemy NPCs, while the worker can be assigned to chop trees (and thus earn words). At the start of the game, one of each villager type is spawned in as a starting population. Each NPC will level up by doing their main tasks, making them more efficient. This is not explicitly expressed to the player but can be discovered naturally through gameplay. 

While this starting population is enough to get started with terraforming tiles and defeating enemies, it is not enough to beat the game. A core part of any god game is growing your population and becoming stronger. We link this mechanic, too, with the generative model. In order to obtain more villager NPCs, the player must terraform maps with ``house'' tiles. Any of these three house-related tiles will spawn a new villager (of a random type) at the start of the next day. This makes house-related words a valuable resource and method of multiplying the players power. 

\subsection{Monsters}
Monsters are the main antagonist of the game. They appear as skeletons and spawn with increasing frequency as the game goes on. Monsters attack any villagers that go near it, presenting a constant threat to the player. Each monsters has much more health compared to an individual villager, and thus they encourage the player to plan carefully and grow their population before attempting to defeat the monsters. The player loses when the monsters take over all the grids on the map.

Monsters are divided into two categories: bosses and minions. Every two minutes, a new boss will spawn and take over a sub-grid on the map. This will continue until the player defeats all the bosses (win) or if the bosses take over all the grids (lose). Boss-occupied grids will continually spawn minion monsters to fight alongside the boss once a villager enters it. The number of minions spawned also scales with game time.

\section{Gameplay}
\begin{figure}
    \centering
    \includegraphics[width=1\linewidth]{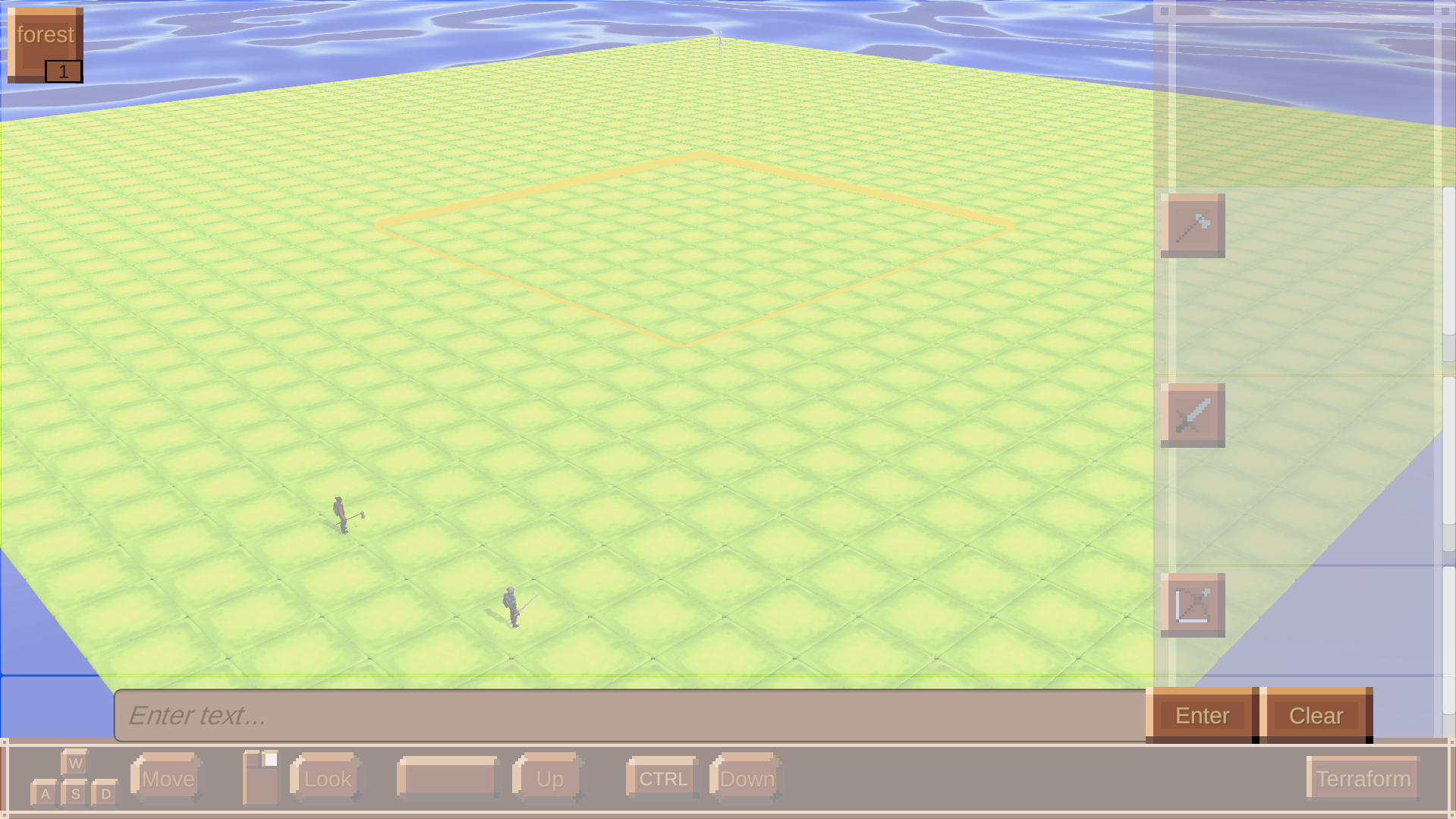}
    \caption{Terraforming Screen: The player can select the words they have collected to modify the terrain.}
    \label{fig:wordbank}
\end{figure}

Each game of God's Innovation Project starts on a large empty field. The only thing the player can see are one villager of each type on the nearest grid and a monster on the furthest one. The player is initially instructed to press the terraform button on their screen. Pressing the button will prompt the player to select one of the grids in front of them. Once selected, the players word bank pops up on screen (Figure \ref{fig:wordbank}), and the player can spawn in their first forest.

From here on the player is free to do whatever they would like. While the optimal goal would be to collect more words via trees and villagers via houses, the open-ended gameplay incentivises exploration. Through experimentation with the terraforming system, the player may encounter special tiles - such as flowers that can heal villagers. These mechanics allow for strategic play, and alternate paths that allow for an easier victory.

To encourage the player's exploration of the prompting system, we do not directly provide details to the player on any of the tile mechanics. Instead, the player has to experiment with different prompts to figure out what they can do and how they can win.



\begin{figure}
    \centering
    \includegraphics[width=0.75\linewidth]{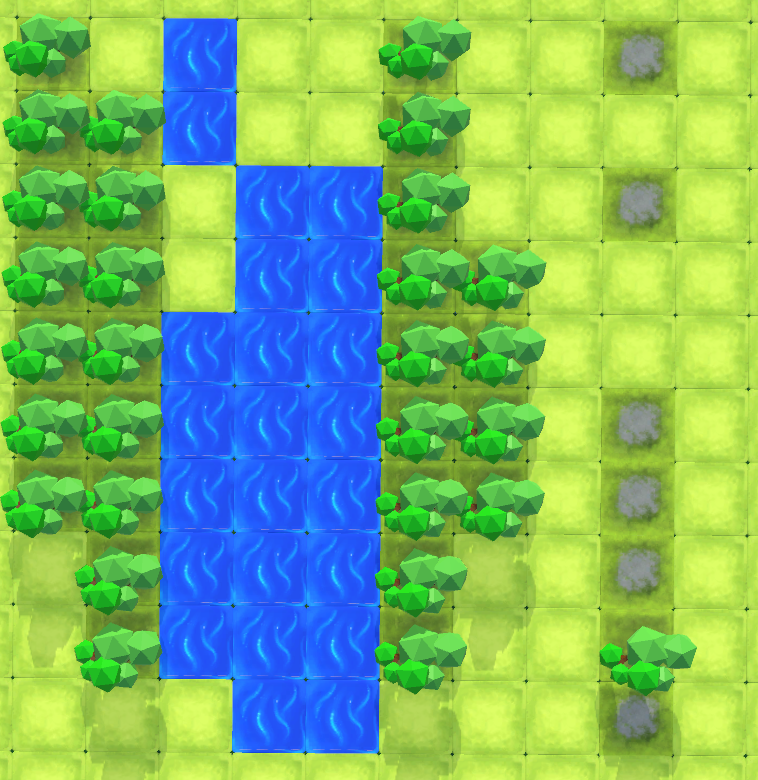}
    \caption{Prompt: a river in a forest}
    \label{fig:river-forest}
\end{figure}
\begin{figure}
    \centering
    \includegraphics[width=0.75\linewidth]{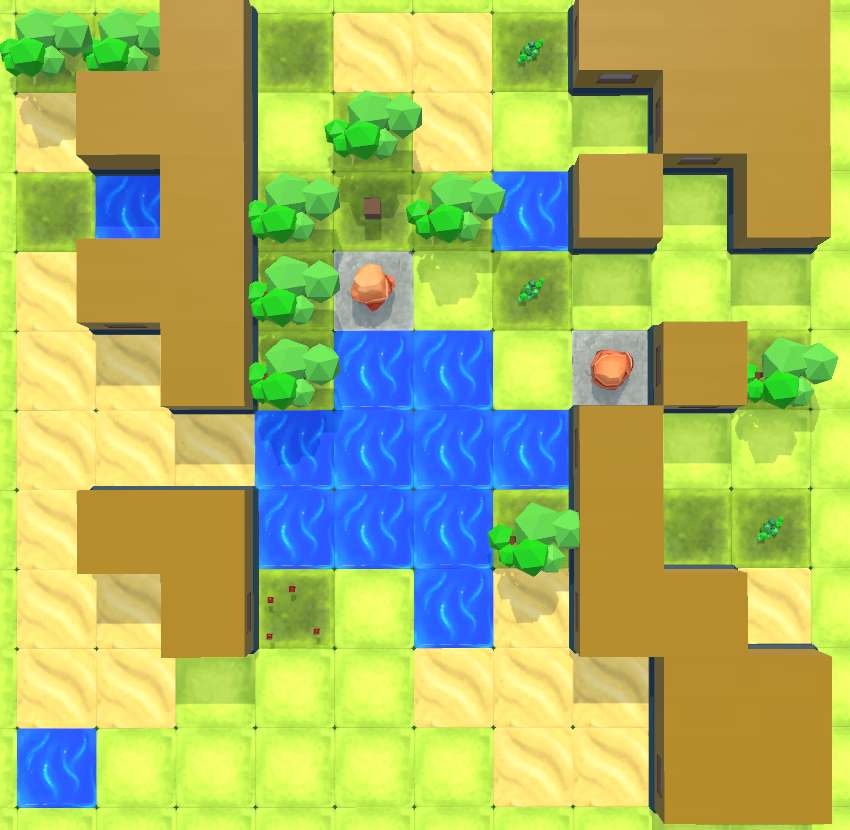}
    \caption{Prompt: a flooded village}
    \label{fig:flood-village}
\end{figure}

Figures \ref{fig:river-forest} and \ref{fig:flood-village} present two examples of the types of terrain the model can generate. Figure \ref{fig:river-forest} shows a water body surrounded by trees, satisfying its prompt. Figure \ref{fig:flood-village} shows a group of houses with water present in the middle. This only partially satisfies the prompt as typically more water is expected in a flood situation.

\section{User Study}
To evaluate the impact of generative AI on gameplay in God’s Innovation Project, we conduct a user study. This study aims to assess how players interacted with AI-generated terrain, the relevance of the generated terrain to their inputs, and the overall effect on gameplay, strategy, and immersion. The study utilized a structured questionnaire to collect both quantitative and qualitative data. The questionnaire contained both 5-point Likert scale questions, as well as some free response exploratory questions.

The survey was shared via social media, and received responses from players with a variety of game experience. Participants were provided with a playable version of the game and asked to engage with the terrain generation mechanic. After playing through the game, they completed the online questionnaire to measure their experience.

\subsection{Questionnaire}
The questionnaire was structured into the following key sections:

\subsubsection{Frequency of Terrain Generation Usage}
Participants indicated how often they used the terrain generation feature during gameplay, selecting from predefined options ranging from \textit{Rarely} to \textit{Always}.

\subsubsection{Relevance of Generated Terrain}
Participants assessed how well the AI-generated terrain matched their input prompts, with ratings from \textit{Least relevant} to \textit{Most relevant}.

\subsubsection{Immersion through Generative AI}
An open-ended question asked whether the AI-driven terrain generation enhanced players’ immersion in the game and why.

\subsubsection{Commonly Encountered Terrain Features}
Players rated the frequency of encountering specific terrain features (e.g., \textit{Grass, Flowers, Bushes, Water bodies}) on a 1-5 scale, from \textit{Never} to \textit{A lot}.

\subsubsection{Encouragement of Creative Thinking}
Participants responded with \textit{Yes} or \textit{No} to whether the AI-driven terrain generation encouraged them to experiment and think creatively.

\subsubsection{Notable Prompts and Accuracy of Terrain Generation}
Three open-ended questions captured player feedback on:
\begin{itemize}
    \item The most interesting prompts they used.
    \item Prompts that resulted in accurate terrain generation.
    \item Prompts where the generated terrain did not align with expectations.
\end{itemize}
To aid the player with these questions, we provide the player with a log file that stores the prompts the player used while playing the game.

\subsubsection{Perceived Impact of Generative AI on the Genre}
An open-ended response section invited players to discuss whether generative AI enhances god games and why.

\subsubsection{Player Experience and Familiarity with Strategy Games}
Players rated their experience with strategy or god games on a 5-point scale.

\subsubsection{Experience with AI-Driven Mechanics in Other Games}
Participants provided examples of other games they had played that incorporated AI-driven mechanics.

\subsubsection{Interest in AI-Generated Mechanics in Future Games}
A multiple-choice question determined whether players would be interested in more games featuring AI-generated mechanics.

\subsection{Data Logging}
We store the prompts sent by the player to AI model in a text file. This allowed us to provide logs for accurate survey responses, as well as analysis on prompt variety and length.

\section{Results and Analysis}
\begin{figure}
    \centering
    \includegraphics[width=0.75\linewidth]{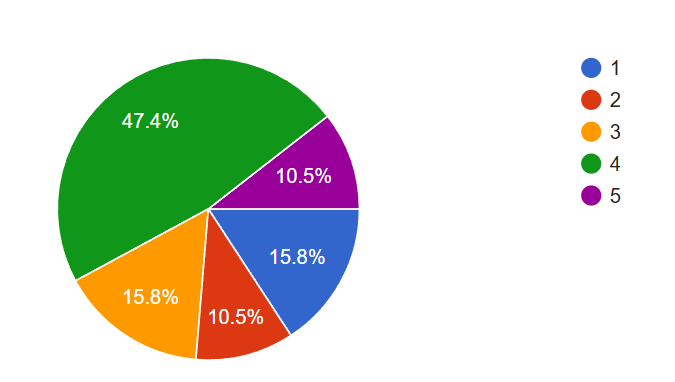}
    \caption{Player Experience With God Games}
    \label{fig:exp}
\end{figure}

We collected results from 19 participants of a varying skill level with god games. Figure \ref{fig:exp} shows that only around half of our respondents considered themselves to be an expert or close to an expert in strategy and god games.

\begin{figure}
    \centering
    \includegraphics[width=0.75\linewidth]{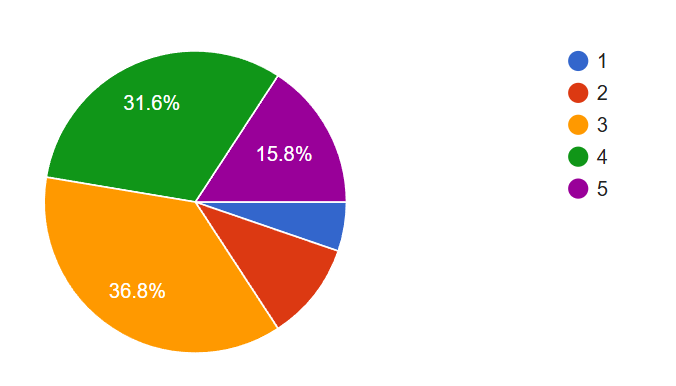}
    \caption{Relevancy of Output Based on Prompt}
    \label{relevancy}
\end{figure}

In figure \ref{relevancy} we can see that on a scale of 1 to 5, 84.2\% players considered the generated terrain to be semantically consistent with the input prompts. Additionally, it provides a valuable human evaluation of a PCGML model. 

\begin{table}[h]
    \centering
    \begin{tabular}{lccccc}
        \textbf{Tile Type} & \textbf{1 (Never) \%} & \textbf{2 \%} & \textbf{3 \%} & \textbf{4 \%} & \textbf{5 (A lot) \%} \\
        Grass             & 5.56  & 11.11  & 11.11  & 5.56  & 66.67 \\
        Flowers           & 11.11 & 33.33  & 22.22  & 16.67 & 16.67 \\
        Bushes            & 5.56  & 38.89  & 16.67  & 22.22 & 16.67 \\
        Path              & 11.11 & 27.78  & 38.89  & 16.67 & 5.56  \\
        Sand              & 5.56  & 11.11  & 50.00  & 22.22 & 11.11 \\
        Rocks/Mountains   & 16.67 & 27.78  & 33.33  & 11.11 & 11.11 \\
        Fences/Posts      & 11.11 & 22.22  & 27.78  & 33.33 & 5.56  \\
        Trees             & 5.56  & 5.56   & 0.00   & 33.33 & 55.56 \\
        Water             & 5.56  & 0.00   & 5.56   & 55.56 & 33.33 \\
        Houses            & 5.56  & 0.00   & 27.78  & 33.33 & 33.33 \\
        Treasure Balls    & 27.78 & 22.22  & 38.89  & 11.11 & 0.00  \\
    \end{tabular}
    \caption{Percentage distribution of tile occurrences as perceived by players.}
    \label{tab:tile_distribution}
\end{table}

In table \ref{tab:tile_distribution} we can observe that most of the tiles appear at a relatively equal frequency. The exception to these rules are \textit{grass, trees, water, }and\textit{ houses}. The \textit{tree} tiles were given some bias as the first word obtained was ''forest'' to improve the player's gaming experience with obtaining prompts to achieve a win state. \textit{Grass} was used as the basic tile in both the game and the model training so it is only natural to see it the most often. The frequency of the house tile can be linked to it being required to spawn more villagers. This can also bee seen in the player logs(Table \ref{tab:tile_prompt_percentage}).

The \textit{water} tile, on the other hand, had an unexpected frequency. Based on player prompt logs as seen in table \ref{tab:tile_prompt_percentage}, we can see that players tended to use water-related prompts at a biased rate compared to other prompts.

\begin{table}[h]
    \centering
    \begin{tabular}{lc}
        \textbf{Tile Type} & \textbf{Percentage (\%)} \\
        Grass & 2.48 \\
        Flowers & 1.98 \\
        Bushes & 0.99 \\
        Path & 3.96 \\
        Sand & 3.96 \\
        Rocks/Mountains & 2.48 \\
        Fences/Posts & 2.97 \\
        Trees & 2.97 \\
        Water & 11.39 \\
        Houses & 9.90 \\
        Treasure Balls & 3.47 \\
    \end{tabular}
    \caption{Percentage distribution of prompts associated with each tile type.}
    \label{tab:tile_prompt_percentage}
\end{table}

On the opposite end, the \textit{treasure ball} item rarely generated. This is expected behavior as those are meant to be rare items that boost the player's playthrough occasionally.

\textit{Fences/posts} and \textit{path} tiles had a mixed but relatively middle of the pack distribution. Similarly \textit{flowers, bushes,} and \textit{sand} had a relatively even distribution meaning that the game is able to provide a varied experience per player.

\begin{figure}
    \centering
    \includegraphics[width=0.75\linewidth]{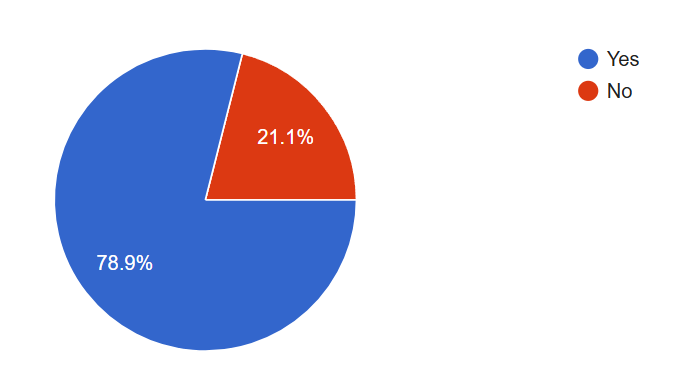}
    \caption{Percentage of players thought outside the box.}
    \label{fig:outsidebox}
\end{figure}

In figure \ref{fig:outsidebox} we can see that 78.9\% of players felt encouraged to think outside the box while playing the game. This fact is further corroborated by the logged prompts. 

\begin{table}[h]
    \centering
    \begin{tabular}{lc}
        \textbf{Prompt Length} & \textbf{Percentage (\%)} \\
        1 Word  & 53.47 \\
        2 Words & 24.75 \\
        3 Words &  7.43 \\
        4 Words &  4.46 \\
        5+ Words &  9.90 \\
    \end{tabular}
    \caption{Percentage distribution of prompt lengths used in the game.}
    \label{tab:prompt_length_distribution}
\end{table}

In table \ref{tab:prompt_length_distribution} we can see that many players were encouraged to try using longer prompts while terraforming. This implies that players were engaging with the game to actively collect words to try out new prompts that they could use to their advantage.

\section{Discussion}
The results of our study provide key insights into how generative AI can serve as an engaging and dynamic game mechanic. Below, we discuss the implications of our findings in relation to player interaction, creativity, and game balance.

\subsection{Engagement and Creativity}
Figure \ref{fig:outsidebox} shows that players were highly engaged with the terraforming mechanic, using it as a tool for strategic and creative problem-solving. The ability to experiment with different prompts and witness immediate changes in the game world provided a sense of agency and discovery, key factors in player immersion.

Furthermore, Table \ref{tab:prompt_length_distribution} indicates that while the majority of player inputs were single-word prompts (53.47\%), a significant number of inputs were multi-word prompts (up to 5+ words at 9.90\%). This implies that players were motivated to collect words in order to form more complex and refined prompts, showcasing an emergent behavior of strategic vocabulary expansion.

With these results, we can assume that using PCGML engages the player in a unique and interesting way. It changes the behavior of the player and forces them to formulate new strategies to succeed in the game. Additionally, from a game design perspective, PCGML creates an interesting gameplay loop that needs an interesting way to generate content for the player using the player's inputs.

\subsection{AI-Generated Content and Player Expectations}
One of the most important aspects of integrating generative AI into a game is ensuring that the output aligns with player expectations. Figure \ref{relevancy} shows that 84.2\% of players found the AI-generated terrain to be relevant to their input. This high level of satisfaction demonstrates the effectiveness of the AI model in translating text prompts into meaningful and coherent terrain.

However, there were instances where the generated terrain did not meet expectations. Based on player feedback, some unexpected outputs stemmed from ambiguous or highly abstract prompts. While this unpredictability can sometimes enhance player experimentation, it can also lead to frustration if players feel they lack sufficient control over the terrain generation process.

In this aspect of PCGML, we can still see further improvement. Real-time PCGML is still in its infancy when compared to traditional PCG. In our game, we had to restrict the prompts and words a player could input into the model. This was to avoid nonsensical prompts that the model could not recognize. As this aspect of game design is further explored and improved on, we will see improvements in the terrain generation as well as reduced in unexpected results as per the player even when given far more freedom.

\subsection{Balancing AI-Generated Content}
Table \ref{tab:tile_distribution} highlights the distribution of different tiles appearing in the game. Notably, certain tiles such as grass, trees, water, and houses appeared more frequently due to their importance in gameplay mechanics. This bias was intentionally introduced to enhance the player's experience, ensuring they had ample opportunities to obtain words and create villagers.

The water tile, however, emerged more frequently than expected, likely due to player preference for water-related prompts. This unintended bias suggests that player behavior can influence the perceived randomness of AI-generated content, an important factor to consider in future AI-driven game mechanics.

On the other hand, the treasure ball item was rarely generated, aligning with its intended role as a rare, high-value game element. This confirms that the AI model can successfully enforce rarity in certain assets, maintaining game balance.

\subsection{Gamifying generative models}
This work represents an early exploration of how generative models can serve as the basis for entire gameplay loops. Current generative models come with drawbacks that make them unnatractive for a commercial product, such as lack of generalization or expensive inference. However, with the current explosive progress in the field of generative AI, generative models may soon become an appealing prospect for game developers.

The Five Dollar Model used in this game suffers from quite a few drawbacks: limited generalization, limited expressivity through a fixed tileset, and nearly deterministic generation. Despite this, players largely enjoyed the AI mechanic and the majority thought that generated maps matched their prompts. We hypothesize that gamifying the text-to-image process was able to avert or hide many of these faults.

By constraining our model via by the word resource system and word pool, we lessen the likelihood of out-of-distribution prompts. Additionally, the game implicitly induces non-nonsensical prompts, providing limited words to use with no regard for grammar or syntax. From the player's perspective, a sub-par generated map results from limited in-game resources, rather than due to an under-performing model.

Further investigation into constrained text-to-image generation and gamification of model prompting is a promising area for future work, and may provide an avenue to create engaging gameplay experiences using imperfect models.

\subsection{Limitation and Future Improvements}
While the study highlights the potential of generative AI in game mechanics, there are several areas for improvement:

\begin{itemize}
    \item \textbf{Prompt Refinement:} Providing players with clearer guidance on how different prompts influence terrain generation could reduce confusion and improve control.
    \item \textbf{Enhanced AI Training:} Expanding the training dataset with more diverse tile arrangements could improve terrain variety.
    \item \textbf{Dynamic Difficulty Adjustment:} AI-generated terrain could be balanced in real time based on player progression, ensuring that players do not create overwhelmingly advantageous or disadvantageous environments.
\end{itemize}

\section{Conclusion}
This paper explored the integration of generative AI as a core game mechanic in God’s Innovation Project. By allowing players to dynamically terraform their environment using AI-generated terrain, we provided a unique gameplay experience that encouraged creativity, experimentation, and strategic thinking.

Through our user study, we found that the generative AI system was well-received, with the majority of players finding the generated terrain relevant to their prompts and feeling encouraged to experiment. We also observed that players actively engaged with the word-collection mechanic to expand their available prompts, reinforcing the idea that AI-driven mechanics can enhance interactivity and immersion.

However, the study also revealed areas for refinement, including better handling of ambiguous prompts and improved balance in AI-generated terrain distribution. Future iterations of this system could explore adaptive AI, where terrain generation evolves based on player behavior, as well as multi-modal AI integration, incorporating additional generative models for further enhancing world-building.

Ultimately, this research demonstrates that generative AI can move beyond traditional applications and become an integral, interactive element in video games. As AI technology continues to evolve, its potential in game design is vast, offering new ways to create dynamic, player-driven experiences.

\end{document}